\documentclass[aps,prl,twocolumn,showpacs,superscriptaddress,groupedaddress,nofootinbib]{revtex4}  

\usepackage{graphicx}  
\usepackage{dcolumn}   
\usepackage{bm}        
\usepackage{amssymb}   
\usepackage{amsmath}
\usepackage{color}

\def\be{\begin{equation}}
\def\ee{\end{equation}}
\def\bea{\begin{eqnarray}}
\def\eea{\end{eqnarray}}

\begin{document}

\widetext
{\flushright{DESY 18-099}}
{\flushright{SI-HEP-2018-20}}
{\flushright{QEFT-2018-12}}

\title{Discovery potential of stable and near-threshold  doubly heavy tetraquarks at the LHC}

\author{Ahmed~Ali}
\email{ahmed.ali@desy.de}
\affiliation{Deutsches Elektronen-Synchrotron DESY, D-22607 Hamburg, Germany}
\author{Qin Qin}
\email{qin@physik.uni-siegen.de}
\affiliation{Theoretische Physik 1, Naturwissenschaftlich-Technische Fakult\"{a}t,
Universit\"{a}t Siegen, Walter-Flex-Strasse 3, D-57068 Siegen, Germany}
\author{Wei Wang}
\email{wei.wang@sjtu.edu.cn}
\affiliation{INPAC, Shanghai Key Laboratory for Particle Physics and Cosmology, MOE Key Laboratory for Particle Physics,
School of Physics and Astronomy, Shanghai Jiao Tong University, Shanghai 200240, China} 
\date{\today}

\begin{abstract}
We study the LHC discovery potential of the double-bottom tetraquarks $bb \bar u \bar d$, $bb \bar u \bar s$ and $bb \bar d \bar s$,
the lightest of which having $J^P=1^+$, called  $T^{\{bb\}}_{[\bar u \bar d]}$,  $T^{\{bb\}}_{[\bar u \bar s]}$ and $T^{\{bb\}}_{[\bar d \bar s]}$, are expected  to be stable against strong decays. Employing the 
Monte Carlo generators 
MadGraph5$\_$aMC@NLO and Pythia6, we simulate the process $pp\to \bar b b\bar b b +X$  and calculate the
$b b$-diquark  jet configurations, specified by the invariant mass interval $M_{bb} < M_{T^{\{bb\}}_{[\bar q \bar q']}} + \Delta M$.
Estimates of $\Delta M$  from the measured product $\sigma(pp \to B_c^+ +X){\cal B}(B_c^+\to J/\psi\pi^+)$ are presented and used to get the
 $b b$-diquark  jet cross sections in double-bottom hadrons  $\sigma(pp \to H_{\{bb\}} +X)$,
 where $H_{\{bb\}} $ represent tetraquarks and baryons. This 
is combined with the LHCb data on the fragmentation $b \to \Lambda_b$ and $b \to B$  to
obtain $\sigma(pp\to T^{\{bb\}}_{[\bar u\bar d]} +X) = (2.8 ^{+1.0}_{-0.7})\ \text{nb} $ at $\sqrt{s}$ = 13 TeV,
and about a quarter of this for the $T^{\{bb\}}_{[\bar{u}\bar{s}]}$ and $T^{\{bb\}}_{[\bar d \bar s]}$, each.
We also present estimates of the production cross sections for the mixed
bottom-charm tetraquarks, $bc \bar u \bar d$, $bc \bar u \bar s$ and $bc \bar d \bar s$, 
obtaining $\sigma(pp\to T^{[bc]}_{[\bar u \bar d]} +X)= (103^{+39}_{-25})\ {\rm nb}$ at $\sqrt{s}$ = 13 TeV, 
and the related ones having  $T^{[bc]}_{[\bar u \bar s]}$ and $T^{[bc]}_{[\bar d \bar s]}$. They have 
excellent discovery potential at the LHC, as their branching ratios in  various charge combinations of 
$BD_{(s)} (\gamma)$ are anticipated to be large.
\end{abstract}

\maketitle

{\it Introduction}:
 The discovery of  $X(3872)$, followed by well over a dozen related mesonic states,   $X$, $Y$, $Z$, and
 two baryonic states~$P_c(4380)$ and  $P_c(4450)$, has opened a second layer of ``extraordinary" hadrons
 in QCD, containing four and five valence quarks and antiquarks~\cite{Tanabashi:2018}. However, their dynamics is not yet deciphered and is under
 intense study. The competing theoretical models put forward  can be roughly classified into two categories:  those
reflecting the residual QCD long-distance effects, dominated by meson exchanges, and those reflecting genuine short-distance
interactions, dominated  by gluon exchanges.  Their spectroscopy,  production and decay characteristics are discussed in a number of  reviews~\cite{Ali:2017jda,Esposito:2016noz,Chen:2016qju,Guo:2017jvc,Olsen:2017bmm}.

Based on the experimental observation  of doubly-charmed baryons~\cite{Aaij:2017ueg} and  Heavy Quark Symmetry (HQS),
recent theoretical insights have brought new 
perspectives, implying that doubly-heavy tetraquarks (DHTQ) $Q_i Q_j \bar q_k \bar q_\ell$ must exist in the HQS limit.
 Here $Q_i, Q_j$ are either $b$ or $c$ quarks, and $\bar q_k,~\bar q_\ell $ are light ($\bar u, \bar d, \bar s$)
antiquarks. The existence of such tetraquarks was already suggested in the earlier 
works~\cite{Ader:1981db, Manohar:1992nd}, but this argument has received a great impetus from proofs based on HQS and
lattice-QCD ~\cite{Karliner:2017qjm, Eichten:2017ffp, Francis:2016hui, Bicudo:2017szl, Junnarkar:2017sey,Mehen:2017nrh, Czarnecki:2017vco}. 
 In particular, HQS relates the DHTQ masses to those of  double-heavy baryons, heavy-light baryons, 
and heavy-light mesons. As the light degrees of freedom in these hadrons are similar, we anticipate that the heavy quark -
heavy diquark symmetry has implications for other non-perturbative aspects as well. In particular, this symmetry can be
used as a quantitative guide in the analysis of the current and anticipated data.

The lightest of the $bb \bar u \bar d$, $bb \bar u \bar s$, and $bb \bar d \bar s$ states are
anticipated to be stable against strong decays. Heavier $bb  \bar q_k \bar q_\ell$ states, as well as the double-charm states
$cc  \bar q_k \bar q_\ell$, and the mixed bottom-charm tetraquark states $bc  \bar q_k \bar q_\ell$, on the other hand,
are estimated to have masses above their respective thresholds. The latter are likely to dissociate into pairs of heavy-light mesons, with large branching ratios, some of which may appear as ``double-flavor"  narrow 
resonances~\cite{Esposito:2013fma,Karliner:2017qjm,Eichten:2017ffp,Luo:2017eub}. 
None of these stable or near-threshold DHTQ mesons has so far been seen experimentally. Observing them would establish the existence of tetraquarks, underscoring the role of diquarks,  with well-defined color and spin quantum numbers
~\cite{Jaffe:1976ig,Jaffe:2003sg,Maiani:2004vq}, as fundamental constituents of hadronic matter.

Our main focus is  to develop the expectations about the production of some of the DHTQ mesons, in particular, the double-bottom $J^P = 1^+$ tetraquarks
$T^{\{bb\}}_{[\bar u \bar d]}$, and the related ones $T^{\{bb\}}_{[\bar u \bar s]}$ and $T^{\{bb\}}_{[\bar d \bar s]}$.
The standard calculational technique,  NRQCD and related frameworks~\cite{Brambilla:2010cs}, however, 
can not be used at present, as the hadronic matrix elements required for tetraquark production are unknown.
The DHTQ decay products are  expected to lie in well-collimated double-heavy-diquark jets, which are formed in high energy collisions.  These configurations can be calculated in perturbative QCD and, combined with non-perturbative (fragmentation) aspects measured in $b$-quark jets, enable us  to estimate the cross sections of interest.

 In a previous paper~\cite{Ali:2018ifm}, we have studied the production of  double-bottom tetraquarks at  a Tera-Z factory in $e^+e^-$ collision, employing the $bb$-diquark jet configurations in which such tetraquarks are likely to be produced.
In this Letter, we study the production of DHTQ states at the LHC, making use of the impressive LHCb data
on $pp \to B_c +X$~\cite{1411.2943} and $b$-hadron production fractions in  pp collisions~\cite{Aaij:2011jp,Aaij:2014jyk}.
 Also, double-bottomonium production has been observed at the LHC, with CMS reporting
a cross section $\sigma(pp \to \Upsilon (1S) \Upsilon(1S) +X)=68 \pm 15$ pb at $\sqrt{s}= 8$ TeV \cite{Khachatryan:2016ydm}. This is the first step in the searches of double-bottom tetraquarks, such as $pp \to T^{\{bb\}}_{[\bar u \bar d]} +X$, as both final states  involve different fragmentation of the same underlying partonic process $pp \to b \bar b b \bar b +X$. Using the
Monte Carlo generators 
MadGraph5$\_$aMC@NLO~\cite{Alwall:2014hca} and Pythia6~\cite{Sjostrand:2006za}, we simulate the process $pp\to \bar b b\bar b b +X$ and estimate that the  production cross section $\sigma(pp \to T^{\{bb\}}_{[\bar u \bar d]} +X)$
 can reach a few nb. 
Replacing a bottom quark by a charm quark, we also simulate the process
 $pp\to b\bar b c\bar c +X$ and  calculate the production of the mixed bottom-charm tetraquarks
 $T^{[bc]}_{[\bar u \bar d]}$,  $T^{[bc]}_{[\bar u \bar s]}$ and $T^{[bc]}_{[\bar d \bar s]}$, having $J^P=0^+$,
 and their $J^P=1^+$ partners, which are estimated to lie above their corresponding heavy-light mesonic thresholds.
  We find that the cross sections for these tetraquarks may reach ${\cal O} (50)~\text{nb}$. As LHCb is projected to collect an integrated luminosity of
  50 fb$^{-1}$ in Runs 1 - 4~\cite{Bediaga:2012py,Bediaga:2012uyd,Carbone:2018}, the prospects  of discovering these tetraquarks are excellent.

{\it Production of double-bottom tetraquarks at the LHC}:
We start by recalling the production and decays of the known doubly-heavy meson $B_c^\pm$ in
the process $pp \to b \bar b c \bar c + X \to B_c^\pm +X$, which serves as the benchmark for our calculations.
At $\sqrt{s}= 8$ TeV,  the LHCb collaboration has measured the ratio~\cite{1411.2943}\footnote{Throughout this
Letter, charge conjugation is assumed.} 
\begin{eqnarray}
R &\equiv& \frac{\sigma(B_c^+){\cal B}(B_c^+\to J/\psi\pi^+)}{\sigma(B^+){\cal B}(B^+\to J/\psi K^+)} \nonumber\\
&=&(0.683\pm0.018\pm0.009)\%, 
\end{eqnarray}
where $0<p_\text{T}<20$ GeV, and $2.0<y<4.5$, with~$p_T$ 
and~$y$ being the component of the momentum transverse to the proton 
beam and rapidity, respectively.

This value is consistent with the previous  LHCb measurement~\cite{Aaij:2012dd}.  
At $\sqrt{s}=7$ TeV, the $B^+$ production cross section is measured as~\cite{1710.04921} 
\begin{eqnarray}
\sigma(B^+) = (43.0\pm0.2\pm2.5\pm1.7)\ \text{$\mu$b}, 
\end{eqnarray}
with the same kinematic cuts. Using MadGraph~\cite{Alwall:2014hca} and Pythia~\cite{Sjostrand:2006za}, we find 
 that the 8 TeV cross section is expected to be enhanced by about 19\%, compared with the 7 TeV cross section,
\footnote{With the cuts $0<p_\text{T}<20$ GeV, and $2.0<y<4.5$ and setting $m_b$ = 4.9 GeV, 
we find that the $b\bar{b}$ cross sections at the 7 and 8 TeV LHC are about 80 $\mu$b and 
95 $\mu$b, respectively. This ratio is less sensitive to the   hadronization. } 
which is consistent with $20\%$ used in~\cite{1411.2943}. Using the above results, and the branching 
ratio~\cite{Tanabashi:2018}
\begin{eqnarray}
{\cal B}(B^+\to J/\psi K^+)=(1.026\pm0.031)\times 10^{-3}, 
\end{eqnarray} 
we find:
\begin{eqnarray}\label{BcXsectionBr}
\sigma(B_c^+){\cal B}(B_c^+\to J/\psi\pi^+)= (0.36\pm0.03)\ \text{nb}. 
\end{eqnarray}

To extract the cross section from the above product, we need to know ${\cal B}(B_c^+\to J/\psi\pi^+)$, which is,
in general, model-dependent. Noting that there is considerable spread in the predicted value of this quantity in the
literature, we use two calculations of the more recent vintage, which we consider more reliable, based on the perturbative QCD approach(pQCD)~\cite{1602.08918},
and on the NLO non-relativistic QCD(NRQCD)~\cite{Qiao:2012hp}, which yield:
\begin{equation}\begin{split}
&{\cal B}(B_c^+\to J/\psi\pi^+) = (2.6^{+0.6+0.2+0.8}_{-0.4-0.2-0.2})\times10^{-3}\ \text{(pQCD)},\\
&{\cal B}(B_c^+\to J/\psi\pi^+) = (2.91^{+0.15+0.40}_{-0.42-0.27})\times10^{-3}\ \text{(NRQCD)}.
\end{split}
\end{equation}
With this, the production cross section  $\sigma(pp \to B_c^+ + X)$ at  $\sqrt{s}=8$ TeV is estimated as:
\begin{equation}\begin{split}
&\sigma(pp\to B_c^+ X) = (139^{+34}_{-41})\ {\rm nb} \ \text{(pQCD)},\\
&\sigma(pp\to B_c^+ X) = (124^{+28}_{-19})\ {\rm nb} \ \text{(NRQCD)}. 
\end{split}
\end{equation}
The implicit model-dependence can be checked by using the  ratios of the semileptonic decays of the
$B_c^\pm$ and $B^\pm$, which have a much larger statistics.

Next, we use  MadGraph~\cite{Alwall:2014hca} to  calculate the cross section for the process
 $pp\to \bar bb\bar cc +X$ at $\sqrt{s}=8$ TeV, which
yields 
\begin{eqnarray}
\sigma(pp\to \bar bb\bar cc +X) = (4.79 \pm 0.08) \times 10^3\ {\rm nb}. 
\end{eqnarray}
As the quarks involved are heavy, charm or bottom, their masses regulate the infrared singularity, which would be
present for the massless quarks. Hence, at the generation level in our simulations, there are no partonic cuts,
i.\,e., in the gluon fusion 
$g g \to b \bar b c \bar c$ and annihilation of the light quark and antiquark 
$q \bar q \to b \bar b c \bar c$. 
This determines for us the fragmentation fraction: 
\begin{equation}\begin{split}
&f(c \bar b \to B_c^+) = (2.9^{+0.7}_{-0.8})\% \ \text{(pQCD)},\\
&f(c \bar b \to B_c^+) = (2.6^{+0.5}_{-0.3})\% \ \text{(NRQCD)}. 
\end{split}\end{equation}
Here, the $B_c^+$ mesons survive the cuts $p_\text{T}<$ 20 GeV and 2.0 $< y <$ 4.5. For the
fragmentation to take place, both the $b$ and the $\bar c$ quarks have to be collinear in a well-collimated jet, defined by an
invariant mass interval $\Delta M$. The fragmentation products include  also the excited states of $B_c^+$, which feed to inclusive $B_c^+$ production through subsequent decays.
We estimate  the value of  $\Delta M$ so as to reproduce the above fragmentation ratio.
This yields:
\begin{equation}\begin{split}
&\Delta M=(2.0^{+0.5}_{-0.4})\ {\rm GeV} \ \text{(pQCD)},\\
&\Delta M=(1.9^{+0.3}_{-0.3})\ {\rm GeV} \ \text{(NRQCD)},
\end{split}\end{equation}
which is consistent with $\Delta M$ = (2.2 - 4.0) GeV that we obtained from simulating the
$Z$ decays in~\cite{Ali:2018ifm}, using NRQCD for $ \sigma(e^+e^- \to B_c +X) $~\cite{Yang:2011ps}, but more
precise. The method of $\Delta M$ determination used here is, however, on a firmer footing as
the experimental measurement of $\sigma(B_c^+)  \times {\cal B }(B_c^+ \to J/\psi \pi^+)$ by LHCb,  
together with the perturbative QCD estimates of the cross section $\sigma(pp \to b \bar b c \bar c +X)$, 
provides the normalization for the inclusive fragmentation $f(c\bar b\to B_c^+)$.

\begin{figure}[tb] 
\begin{center}
\includegraphics[width=0.4\textwidth]{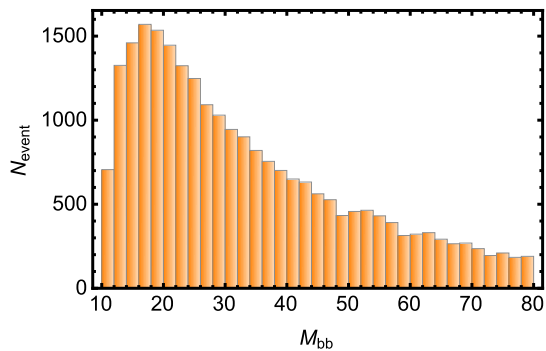} 
\end{center}
\caption{The $bb$-quark-pair invariant-mass distribution for the process $pp \to (bb)_{\rm jet} + \bar{b} + \bar{b} +X$  at $\sqrt{s}=13$ TeV, obtained by generating $10^{5}$ events using MadGraph and Pythia6 at the  NLO accuracy. 
} \label{fig:invmass} 
\end{figure} 

\begin{figure}[ht!]
\centering
\includegraphics[width=70mm]{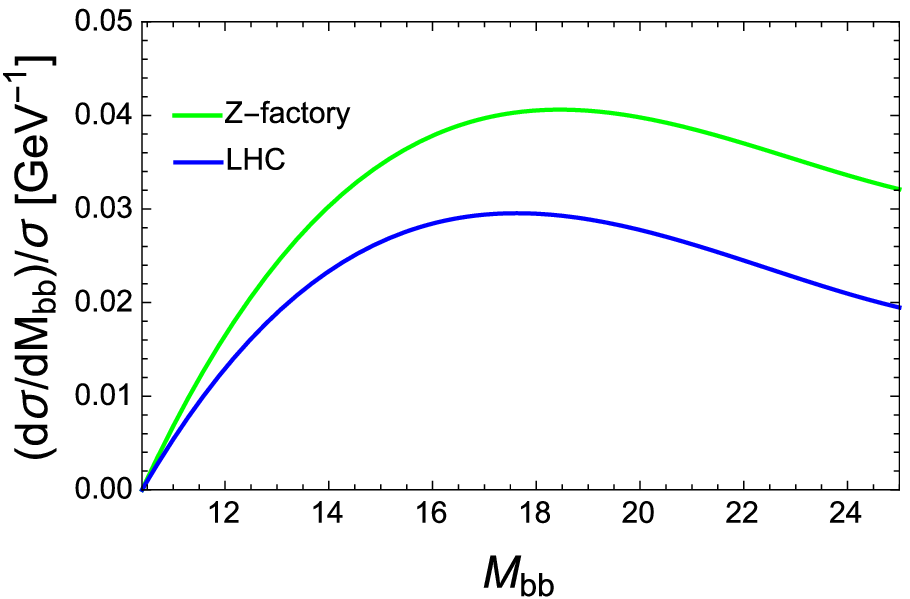} 
  \hspace{0.2in}
\includegraphics[width=70mm]{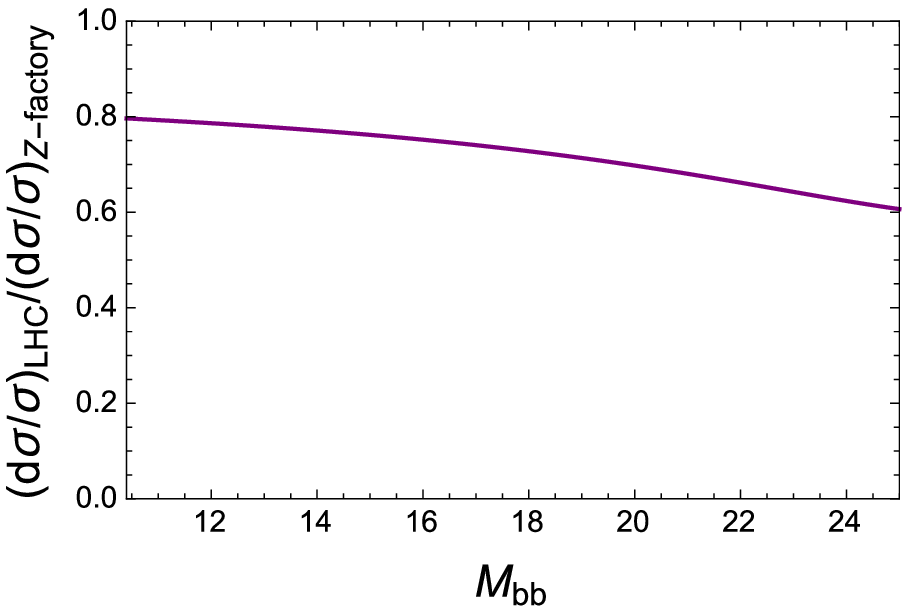}
\caption{Normalized differential cross section $\frac{1}{\sigma} [d\sigma (pp \to (bb)_{\rm jet} + \bar{b} + \bar{b} +X)/d M_{bb}]$  at the LHC (13 TeV) versus  the corresponding cross section at the $Z$ pole  (upper panel) and the ratio of the normalized differential cross sections  (lower panel). See text for details.}
\label{fig:ZfactoryLHC}
\end{figure}

For $pp\to b \bar b b \bar b +X$, 
we have  generated $10^5$ showered events at $\sqrt{s}=13$ TeV with MadGraph~\cite{Alwall:2014hca} and Pythia6~\cite{Sjostrand:2006za} at the NLO accuracy.   
The cross section $\sigma(pp\to b \bar b b \bar b +X)$, involving the $gg$ and $q\bar{q}$ partons, is evaluated by MadGraph to be (463 $\pm$ 4) nb. We also find that the contribution from the $Z$-induced 
processes, $ (pp \to Z +X \to b \bar b b \bar b +X)$ and $ (pp \to Z b \bar b + X\to b \bar b b \bar b +X)$, is down by three orders of magnitude, and hence is not considered any further.

\begin{figure}[tb] 
\begin{center}
\includegraphics[width=0.4\textwidth]{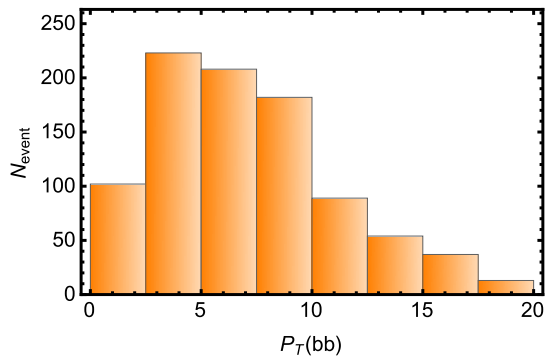} 
\end{center}
\caption{The $bb$-quark-pair $P_\text{T}$ distribution for the process $pp \to (bb)_{\rm jet} + \bar{b} + \bar{b} +X$ 
at $\sqrt{s}=13$ TeV, obtained by generating $10^{5}$ events using MadGraph and Pythia6 at the  NLO accuracy. The $ (bb)_{\rm jet}$ is defined by the interval $M _{bb} (\Delta M)$ with $\Delta M=2.0$ GeV. 
} \label{fig:pt} 
\end{figure}

The $b$-quark pair invariant mass distribution  is displayed in Fig.~\ref{fig:invmass}.  We compare the normalized $bb$-invariant mass distribution at the LHC 
($\sqrt{s}=13$ TeV) with the
corresponding  one in $e^+e^-$ collision at the $Z$ pole in Fig.~\ref{fig:ZfactoryLHC},  upper panel, while the lower panel shows the ratio of the two. From this figure, we see that the jet- shapes (normalized distributions) are similar in the
two cases in the  small invariant mass region. Thus, the same 
jet-resolution  criterion can be used in the two processes to estimate the fraction of the $bb$-invariant 
mass in which the $bb$-diquark is likely to fragment into double-bottom hadrons.  We use
$\Delta M=(2.0^{+0.5}_{-0.4})$ GeV, obtained from the analysis of the data on $\sigma(pp \to B_c^+ + X)$ at  $\sqrt{s}=8$ TeV, 
discussed earlier, which yields the following fragmentation fraction and the corresponding cross section 
\begin{eqnarray}
f (b b \to H_{\{bb\}}) =  (3.2^{+1.2}_{-0.8})\%, \\
\sigma(pp\to H_{\{bb\}} +X)= (14.8^{+5.4}_{-3.7})\ {\rm nb}.
\label{eq:Hbb}  
\end{eqnarray}
The double-bottom hadrons~$H_{\{bb\}}$ include the double-heavy tetraquarks $T^{\{bb\}}_{[\bar q \bar q^\prime]}$ 
and the double-bottom baryons $\Xi_{bb}^0 (bbu)$, $\Xi_{bb}^- (bbd)$, and $\Omega_{bb}^- (bbs)$, and their excited
states, as well as some non-resonant open-bottom background, such as $BB^{(*)}+X$, which we expect to be small
due to the stringent  $\Delta M$ cut. The relative fractions of $H_{\{bb\}} \to T^{\{bb\}}_{[\bar q \bar q^\prime]} $ and
  $H_{\{bb\}} \to \Xi_{bb}^0 (bbu)$, $\Xi_{bb}^- (bbd), \Omega_{bb}^- (bbs) $ are not known. In the fragmentation language, they
involve the vacuum excitation of a light anti-diquark pair ($ \bar q \bar q^\prime $) in the former, and of 
 a light quark-antiquark pair in the latter.
  We assume, appealing to the heavy quark - heavy diquark symmetry,  that they are similar to the measured
ones in a single $b$-quark jet, for which LHCb has reported the following $p_T$-dependent ratio~\cite{Aaij:2011jp}:
 \begin{equation}\begin{split}\label{eq:frag}
\left[\frac{f_{\Lambda_b}}{f_{B_u} + f_{B_d}}\right]&(p_\text{T}) = (0.404 \pm 0.036)\\
&\times[1-(0.031\pm0.005)p_\text{T}\text{(GeV)}], 
\end{split}\end{equation}
where we have added in quadrature the various errors quoted in \cite{Aaij:2011jp}.
To use this input, we need to first calculate the $p_\text{T}$-distribution of the $bb$-diquark jet
in $pp \to (bb)_{\rm jet} + \bar{b} + \bar{b} +X$. This is shown in Fig.~\ref{fig:pt}, where the $ (bb)_{\rm jet}$ is defined by the interval $M _{bb} (\Delta M)$ with $\Delta M=2.0$ GeV. 
\begin{figure}[tb] 
\begin{center}
\includegraphics[width=0.4\textwidth]{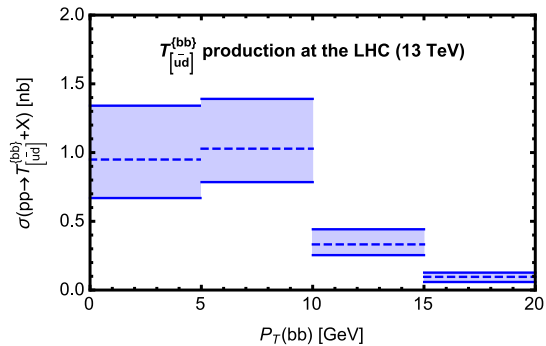} 
\end{center}
\caption{Projected $p_\text{T}$-dependence of tetraquark production cross section
in $pp \to T^{\{bb\}}_{[\bar u\bar d]} +X$ at the LHC for $\sqrt{s}=13$ TeV.
} \label{fig:tetraquark_pT} 
\end{figure} 

We convolute  this distribution with the one measured by LHCb for $\left[\frac{f_{\Lambda_b}}{f_{B_u} + f_{B_d}}\right](p_\text{T})$, given above, and estimate
 the ratio of the $T^{\{bb\}}_{[\bar u\bar d]}$ 
production cross section to the $H_{\{bb\}}$ production cross section:
\begin{eqnarray}
\frac{\sigma(pp\to T^{\{bb\}}_{[\bar u\bar d]} +X)}{\sigma(pp\to H_{\{bb\}}+X)} = 0.195 \pm 0.014,
\end{eqnarray}
with both $H_{\{bb\}}$ and $T^{\{bb\}}_{[\bar u\bar d]}$ having $p_\text{T} <$ 20 GeV. 
Here, we have assumed that the final state $H_{\{bb\}}+X $ is saturated by the
double-bottom baryons and double-bottom tetraquarks, and the ratio of the production rates of the strange and non-strange $B$ mesons, 
$f_s/f_d = 0.256\pm0.020$ measured by the LHCb collaboration \cite{Aaij:2013qqa}, 
also holds in the fragmentation of a $bb$ jet. 
This leads finally to the integrated cross section:
\begin{eqnarray}
\sigma(pp\to T^{\{bb\}}_{[\bar u\bar d]} +X) = (2.8 ^{+1.0}_{-0.7})\ \text{nb}. 
\label{eq:Tbb} 
\end{eqnarray}
The $p_\text{T}$-distribution of the  differential cross section for $pp\to T^{\{bb\}}_{[\bar u\bar d]} + X$  is shown in Fig.~\ref{fig:tetraquark_pT}. 

The double-bottom tetraquark states $ T^{\{bb\}}_{[\bar q\bar q^\prime]}$  worked out in our  paper have isospin zero, and they are the main objects  of interest due to their anticipated stability against strong interactions. They decay weakly, and their search strategies are based on their long lifetimes and weak decay products. The isospin-1 partners of $T^{\{bb\}}_{[\bar u\bar d]} $  may also be present in the jet-cone defined by  $M _{bb} (\Delta M)$, though they involve the so-called "bad" light diquarks, having spin 1. 
  Denoted  as $\{ bb\}  \{\bar{q}_k \bar{q}_l \}$, their masses, $J^P$ quantum numbers, and dominant decay modes have been worked out in the literature, in particular, in \cite{Eichten:2017ffp}. They
 have $J^P=0^+, 1^+, 2^+$, and masses above their corresponding $B^- B^{(*)0 }$ thresholds, with the estimated $Q$ values of
 115, 78, and 136 MeV, respectively. Such states lie typically 200 MeV above the mass of the spin-0 $T^{\{bb\}}_{[\bar u\bar d]} $ state. Thus, apart from their indicated decays, they may also feed down to the isosinglet $T^{\{bb\}}_{[\bar u\bar d]} $ state
 via a pion emission, if allowed by parity conservation.
 Their existence as bound states has, however, not been demonstrated in a lattice based calculations, though in the
 HQS limit, they are expected to be produced at a comparable rate to their isospin-0 partner. Thus, there is at present
 an uncertainty in the composition of $ f (b b \to H_{\{bb\}})$ in Eq. (\ref{eq:Hbb}), reducing the cross section given in Eq. (\ref{eq:Tbb}).   
  
The production cross sections for the double-bottom baryons (summed over the states) are estimated as
\begin{equation}
\sigma(pp \to (\Xi_{bb}^0, \Xi_{bb}^-, \Omega_{bb}^-) + X) : 
\sigma(pp \to  T^{\{bb\}}_{[\bar q \bar q']} + X) \approx 2.4.
\label{eq:BRXibb}
\end{equation}
Thus, we anticipate about twice as many double-bottom baryons as the double-bottom tetraquarks at the 13 TeV LHC. 

The LHCb collaboration is expected to collect about 50 fb$^{-1}$ of data in
 Runs 1-4 ~\cite{Bediaga:2012py,Bediaga:2012uyd,Carbone:2018}, which
would translate into  ${\cal O}(10^8)$ $T^{\{bb\}}_{[\bar u\bar d]}$ events. Taking into account the $s\bar{s}$-suppression,
compared to $d \bar{d}$ or  $u \bar{u}$, we expect approximately a quarter of this number for the other two doubly-bottom
tetraquarks $T^{\{bb\}}_{[\bar u\bar s]}$ and $T^{\{bb\}}_{[\bar s\bar d]}$, each. Their lifetimes are expected to be very similar,
and estimated as 0.8 ps~\cite{Ali:2018ifm}. Their anticipated discovery modes have typical branching ratios of
${\cal O}(10^{-6})$~\cite{Ali:2018ifm,Xing:2018bqt}, much smaller than the one for decays of doubly charmed baryons~\cite{Wang:2017mqp,Wang:2017azm} which are recently observed by LHCb~\cite{Aaij:2017ueg}.  It indicates   that dedicated searches at the LHC will be required to discover them.

{\it Production of $T^{[bc]}_{[\bar u \bar d]}$ at the LHC}:
As already noted, LHCb has collected an impressive amount of $B_c$ events, 
with $2.1\times 10^3$ 
$B_c \to J/\psi \pi$ candidates 
in  2 fb$^{-1}$ pp collisions at 8 TeV~\cite{1411.2943}. As the underlying partonic process is
the same for the tetraquark $T^{[bc]}_{[\bar u \bar d]}$ production, but non-perturbative aspects differ, we evaluate the
production cross section $\sigma(pp \to  T^{[bc]}_{[\bar u \bar d]} +X)$. For that
we generate $10^4$ showered $pp\to b \bar b c \bar c +X$ events at the $pp$  centre-of-mass energy 8 TeV,
using  the generators MadGraph~\cite{Alwall:2014hca} and Pythia6~\cite{Sjostrand:2006za} at the NLO accuracy. 
The cross section  $\sigma(pp\to b \bar b c \bar c +X)$ is evaluated by MadGraph to be $(4.79\pm0.08)$~$\mu$b, which on using  $\Delta M=(2.0^{+0.5}_{-0.4})$ GeV yields the following fragmentation fraction and the corresponding cross section:
\begin{eqnarray}
f (b c \to H_{[bc]}) =  (5.7^{+2.4}_{-1.4})\%, \\
\sigma(pp\to H_{[bc]} +X)= (273^{+113}_{-~66})\ {\rm nb}.
\label{eq:Hbc8}  
\end{eqnarray}
Combined with Eq.~\eqref{eq:frag} for the fragmentation fraction, we get at $\sqrt{s}=8$ TeV
\begin{eqnarray}
\sigma(pp\to T^{[bc]}_{[\bar u \bar d]} +X)= (57^{+22}_{-14})\ {\rm nb},
\label{eq:Hbc8}  
\end{eqnarray}
with $p_\text{T}(T^{[bc]}_{[\bar u \bar d]})<$ 20 GeV. This cross section is larger than an earlier
 estimate~\cite{Yuqi:2011gm}, in which the diquark with heavy quarks is first produced and then fragment into the
  tetraquark state. 
Assuming a detection efficiency of $10^{-6}$, we anticipate $O(10^3)$ $ T^{[bc]}_{[\bar u \bar d]} $ candidate events in
the currently available LHCb data set,
and approximately a quarter of this number for the related tetraquarks $ T^{[bc]}_{[\bar u \bar s]} $ and 
$ T^{[bc]}_{[\bar d \bar s]} $. There is considerable uncertainty in these estimates as the mixed bottom-charm tetraquarks,
as opposed to the stable $bb$-tetraquarks, have $J^P=0^+$, and  $J^P=1^+$, and their relative production rates in the fragmentation of a $cb$-diquark is an additional unknown parameter. They apply to the sum of both the $J^P$ states.
The mass of $T^{[bc]}_{[\bar u \bar d]}$ is estimated in~Ref.~\cite{Eichten:2017ffp} to be 7229 MeV, 
some 83 MeV above the $BD$ threshold, and one expects a narrow resonance in this channel. The masses of the
other two tetraquarks with an $s$-quark, are pitched at 7406 MeV, some 170 MeV above the $B_s D$ threshold~\cite{Eichten:2017ffp},  considerably broadening the resonances. 

Finally, the cross section $\sigma(pp\to b \bar b c \bar c +X)$ at 13 TeV
is evaluated by MadGraph to be  $(8.76 \pm 0.19)$ $\mu$b.
Repeating the steps indicated for the 8 TeV case, we estimate that
\begin{eqnarray}
\sigma(pp\to T^{[bc]}_{[\bar u \bar d]} + X)= (103^{+39}_{-25})\ {\rm nb},
\label{eq:Hbc8}  
\end{eqnarray}
with $p_\text{T}(T^{[bc]}_{[\bar u \bar d]})<$ 20 GeV.

With the LHCb integrated luminosity of 50 fb$^{-1}$, to be reached in Runs 1-4, well over $10^{9}$ $T^{[bc]}_{[\bar u \bar d]}  +X$ events will be produced.  

Finally, we have not considered the double-charm tetraquark states, as their existence as bound states is not established
by the current lattice calculations \cite{Cheung:2017tnt,Guerrieri:2014nxa}.

We would like to thank  Estia Eichten, Tim Gershon,  Marek Karliner, Luciano Maiani, Alexander Parkhomenko, Antonello Polosa,  Gerrit Schierholz, Sheldon Stone, Cen Zhang and Zhi-Jie Zhao for helpful discussions.   This work is supported in part by the National Natural Science Foundation of China under Grant Nos. 11575110, 11655002, 11735010,  the Natural Science Foundation of Shanghai under Grant No.~15DZ2272100, and  the DFG Forschergruppe FOR 1873 ``Quark Flavour Physics and Effective Field Theories".

\end{document}